\documentclass[12pt]{elsarticle}

\usepackage{natbib}
\usepackage[utf8x]{inputenc}
\usepackage[super]{nth}
\usepackage[margin=3cm]{geometry}
\usepackage[hyphens]{url}
\usepackage{hyperref}

\usepackage{multirow}
\usepackage{booktabs} 
\usepackage{graphicx}
\usepackage{subcaption}
\usepackage{comment}
\usepackage{xspace}
\usepackage{color}
\usepackage{flushend}
\usepackage[absolute,overlay]{textpos}

\newcommand{\monroe}{MONROE\xspace}
\newcommand{\TOOL}{ERRANT\xspace}
\newcommand{\TOOLNAME}{EmulatoR of Radio Access NeTworks\xspace}
\newcommand{\HTTP}{\textit{HTTP}\xspace}
\newcommand{\META}{\textit{MetaData}\xspace}
\newcommand{\SPEED}{\textit{SpeedTest}\xspace}
\newcommand{\SIMPLE}{\textit{Simple Emulator}\xspace}

\usepackage[pscoord]{eso-pic}

\begin{document}

\title{\TOOL: Realistic Emulation of Radio Access Networks}

\author{
    Martino Trevisan$^\ddagger$, 
    Ali Safari Khatouni$^\dagger$, 
    Danilo Giordano $^\ddagger$ 
    \\
    \small{$^\dagger$Western University,} \texttt{asafarik@uwo.ca}
    \\
    \small{$^\ddagger$Politecnico di Torino,} \texttt{first.last@polito.it} 
}

\TPshowboxestrue
\TPMargin{0.3cm}
\begin{textblock*}{15.5cm}(3cm,0.8cm)
\footnotesize
\bf
\definecolor{myRed}{rgb}{0.55,0,0}
\color{myRed}
\noindent
Please cite this article as: Martino Trevisan, Ali Safari Khatouni, Danilo Giordano. ERRANT: Realistic Emulation of Radio Access Networks. Computer Networks (2020). DOI: \url{https://doi.org/10.1016/j.comnet.2020.107289}
\end{textblock*}

\makeatletter
\def\ps@pprintTitle{%
 \let\@oddhead\@empty
 \let\@evenhead\@empty
 \def\@oddfoot{}%
 \let\@evenfoot\@oddfoot}

\newcommand\footnoteref[1]{\protected@xdef\@thefnmark{\ref{#1}}\@footnotemark}
\makeatother

\begin{abstract}
Mobile devices drastically changed how people use the Internet. We use smartphones to access a heterogeneous catalog of web services such as news, social networks, audio/video streaming. Differently from wired connections, mobile networks do not offer the same kind of performance stability yet. Thus, service providers have to handle different network scenarios, e.g., 3G or 4G, while promising good end-users' quality of experience (QoE). To ensure that QoE is adequate, it is necessary to thoroughly test applications with a wide range of possible network conditions. 
For this, network emulation is of vital importance as it allows a tester to run experiments with a wide range of network conditions. However, when it comes to mobile networks, the variety of technical characteristics, coupled with the opaque network configurations, makes realistic emulation a challenging task. Most of the freely available emulation tools rely on a simple emulation, offering limited variability performances for each network condition.

In this paper, we propose \TOOL, \TOOLNAME, an open-source tool that emulates mobile networks with a high level of realism, following a data-driven approach. We use a large-scale dataset composed of $100$ k speed test measurements collected from $4$ network operators in $2$ countries. We create $32$ different network profiles based on different countries, operators, radio access technologies, and signal qualities.
For each profile, we obtain both typical behavior and variability for latency, download and upload bandwidth. We use the profiles to create models by means of the Kernel Density Estimation.
Then, \TOOL employs the \texttt{tc-netem} Linux tool and the models for emulation. In this way, \TOOL offers realistic network emulation, in which both \emph{typical} behavior and network variability are accurately recreated.

We validate \TOOL models with an independent dataset of HTTP downloads performed on the same mobile networks as of the profiles. Results show the effectiveness of \TOOL in the emulation of real mobile networks in terms of average behavior and obtained variability. We also show the limitations of a simple emulation, and of other freeware approaches versus \TOOL. Finally, we show two practical use cases to demonstrate the benefits of a dynamic emulation in understanding the performance of web browsing and video streaming. To run new measurement campaigns and create new models, we provide guidelines along with the required open-source code.
\end{abstract}

\maketitle

\section{Introduction}
\label{sec:intro}
Mobile devices dramatically revolutionized the way we interact with the Internet, nowadays offering access with a capacity comparable with the wired network~\cite{vassio2018you}. Therefore, we use mobile devices to access a heterogeneous catalog of services: surf the web to read the news, organize video conferences, watch videos in high quality, and recently, play video games in streaming. 
All these services have different network requirements: a low-quality connection works well to read the news, but it may be insufficient to watch a video or to make a video conference. As such, sensitive services tend to adapt their quality based on the underlying connectivity. For example, modern on-demand video services rely on the so-called \emph{dynamic adaptive streaming}, that adjusts dynamically video bitrate based on network conditions.

To evaluate whether web services work properly with different network conditions a thoroughly testing phase is necessary. As such, over the recent years several tools~\cite{network_link_conditioner, atc,netemul1,netemul2} emerged to evaluate how web services work under different conditions. 
These tools typically \textit{emulate} specific network technologies (e.g., ADSL, 3G, etc.) based on \textit{profiles} defined by different parameters. 
While some commercial tools offer several parameters to tune the connection performance e.g., packet loss, duplicates, out of order packets, link congestion, etc., which could be difficult to tune, all commercial and freeware tools share three key parameters: the \textit{download bandwidth}, the \textit{upload bandwidth}, and the \textit{latency}. 
These parameters are easier to tune, as little bandwidth as a clear impact on the connection, preventing the user to watch a video, while high latency could make video or audio conference unfeasible. 
Based on these three parameters, each tool exposes emulation profiles without providing any information about how such profiles are generated. 
For instance, while the 3G profile in the Google Chrome Developer Tool offers 750 kbps as download bandwidth, the same parameter in the Android Emulator reaches 14 Mbps. 
Moreover, these parameters are imposed without any variability, preventing the tool to emulate the intrinsic fluctuations of mobile networks.

To overcome these limitations using an open-source solution, we propose \TOOL, \TOOLNAME, an emulation tool based on a data-driven approach. \TOOL builds \textit{emulation models} using a large-scale dataset collected within real operational mobile networks by using the mobile nodes of the \monroe platform~\cite{khatouni2017speedtest,midoglu2018monroe}. 
This dataset includes more than $100$ k speed test measurements from $20$ nodes/locations equipped with SIM cards of $4$ operators, located in $2$ distinct countries. 
The dataset is enriched by physical-layer metadata such as  Radio Access Technology (RAT), signal strength, and country to augment the initial dataset with contextual information. 
We use all the available data (i.e., speed test and metadata) to create $32$, different emulation models. 
The first $26$ models describe the mobile network performance for a specific: a network operator, country, RAT (i.e., 3G and 4G), and signal quality. Plus, we provide $6$ additional models to be used as general models describing an operator offering 3G or 4G connectivity with different signal qualities. 
Given that \TOOL is open-source, users willing to contribute can perform their own measurement campaigns to create new emulation models for specific operators, locations, new technology, i.e., 5G.  

\TOOL provides dynamic models describing not only the \emph{typical} (and simple) behavior of mobile networks, but also their variability. 
For this, we employ the Kernel Density Estimation (KDE)~\cite{parzen1962estimation} technique to create accurate models of the observed network, based on the speed test measurements. 
Then, \TOOL emulates a dynamic behavior by extracting the network emulation parameters from the models. 
By extracting new parameters at each run (or changing parameters within a run following the tester decisions) \TOOL can correctly emulate the variability of the network conditions observed in the real networks. 
In other words, our tool allows the user to run \emph{several} experiments, sampling realistic values from the modeled distribution, and, in turn, to observe the distribution of the desired effects. To apply the emulation parameters, under the hood \TOOL uses the \texttt{tc-netem} Linux emulation tool which is the \emph{de facto} standard.

We validate the models comparing the download speed in our emulated environment with those obtained with real-world experiments in which mobile nodes download large-sized HTTP objects. In addition, we compare our results with other free emulators. 
The results show that not only \TOOL models reproduce the \emph{typical} behavior of mobile networks, but also it introduces similar levels of variability which are not present with the other emulators. As such, using \TOOL, a tester can accurately try her service as if she was under quasi-real mobile network conditions. 

Finally, we show how our models help in evaluating the Quality of Experience (QoE) of a user accessing web services. 
We compare the performance of web pages accessed under different network conditions and notice a strong impact on page load time, in terms of absolute values and variability. 
Indeed, simple emulation does not pinpoint the variability QoE-related metrics that would occur in real mobile networks. 
We also show that video streaming as another use case, in which emulating variable network conditions lead to considerably different results in terms of QoE.

We make \TOOL open source and available to the community~\cite{mne} to allow practitioners to test their applications and researchers to reproduce the results. 
This paper extends our preliminary work~\cite{ali2019driver}, where we already have shown the potentialities and limitations of using a data-driven emulator. 
We expanded our work from many points of view:
(i) we now rely on a large-scale speed test measurement campaign;
(ii) we provide a fully-fledged network emulator that models not only latency but also download and upload bandwidth;
(iii) we show practical use cases to illustrate the benefits of our proposal in understanding the behavior of user applications with a particular focus on web browsing and video streaming;
(iv) we provide an open source tool to emulate mobile networks in different operators, locations, and technologies. 
The remainder of this paper is organized as follows. 
Section~\ref{sec:relatedwork} discusses related work. 
Section~\ref{sec:measurement} presents the measurement setup, the datasets, and how to replicate a measurement campaign. 
The model creation and emulation methodology are presented in Section~\ref{sec:methodology}. Section~\ref{sec:emulation_resuts} presents the experimental results, while in Section~\ref{sec:web_use_case} we show two practical use cases.
Finally, Section~\ref{sec:conclusions} concludes the paper. 
\section{Related Work}
\label{sec:relatedwork}

\begin{table*}[]
    \scriptsize
    \begin{center}
    \caption{Emulation parameters of default profiles for 6 freeware testing tools, separately for download (D), upload (U), and latency (L). We omit \emph{bps} for download and upload, and \emph{ms} for latency for the sake of space.}

    \setlength{\tabcolsep}{1.6pt}
    \begin{tabular} {rllllllllllll}
     \toprule
     \multicolumn{1}{c}{\multirow{2}{*}{}} & \multicolumn{3}{c}{\bf 3G} & \multicolumn{3}{c}{\bf 3G Slow} & \multicolumn{3}{c}{\bf 3G Fast} & \multicolumn{3}{c}{\bf  4G}  \\
     \cmidrule(lr){2-4} \cmidrule(lr){5-7} \cmidrule(lr){8-10} \cmidrule(lr){11-13}
     \multicolumn{1}{c}{}                         & D        & U        & L     & D          & U         & L       & D          & U         & L       & D       & U       & L \\ 
    \midrule
     Chrome~\cite{chrome_dev_tools} & 750 k & 250 k & 100 & - & - & - & 1 M & 750 k & 40 & 4 M & 3 M & 20 \\
     WebPageTest~\cite{webpagetest} & 1.6 M & 768 k & 300 & 400 k & 400 k &  400 & 1.6 M  & 768 k & 150 & 12 M & 12 M & 70 \\ \
    BrowserTime~\cite{browsertime_original} & 1.6 M &  768 k &  300 & 780 k & 330 k  & 200 & 1.6 M  & 768 k & 150 & - & - & -  \\
    ATC~\cite{atc}  & 780 k & 330 k  & 200  & 850 k & 420 k & 190 & - & - & - & - & - & - \\  
    Android Emulator~\cite{android_emulator} & 14 M & 5.76 M & 0 & 384 k & 384 k & 35-200 & - & - & - & 173 M & 58 M & 0 \\
    NLC~\cite{network_link_conditioner} & 780 k & 330 k & 100 & - & - & - & - & - & - & 51.2 M & 10.24 M & 65 \\
    \bottomrule
    \end{tabular}
    \label{tab:available_tools}
    \end{center}
\end{table*}

Network monitoring provides crucial information about the performance of operational networks and helps network managers, researchers, and regulators to understand the behavior of the network and users' QoE. 
There are nationwide~\cite{fcc2013,samknows} and international~\cite{alay2017mobicom} efforts to monitor and measure the performance of networks. Non-stop measurements become more important when it comes to the latency and throughput of mobile networks. 
The big challenge is to translate the performance of the network into the users' QoE~\cite{aberdeen_study,bing_google_study}. 

There is a large body of research works that shows the effect of different scenarios in the performance of mobile networks. 
The latency is a critical metric for many applications, e.g., audio streaming applications. Mobile access networks and mobility scenarios present very high variability in performance -- due to the more random nature and complexity of mobile networks. 
A survey of end-to-end delay prediction methods is offered in~\cite{yang2004predicting}. They show several methodologies to estimate latency, e.g., \textit{Queuing Network Modelling}, \textit{System Identification}, \textit{Time Series Approach}, and \textit{Neural Networks}. Nikravesh~\emph{et al.}~\cite{10.1007/978-3-319-04918-2_2} indicate significant performance differences across operators, access technologies, geographic regions and over time; however, they highlight that these variations themselves are not uniform, making network performance difficult to predict. 
Safari~\emph{et al.}~\cite{4g-prediction} illustrate accurate mobile network latency prediction is not effective by using the common machine learning algorithms.   
Mandalari~\emph{et al.}~\cite{Mandalari:roaming} present the impact of roaming on latency with $\approx$ 60 ms or more, depending on geographical distance. 
The feasibility of the latency-sensitive application, teleoperated driving is demonstrated by means of measurements using mobile networks however it could be challenging in some cases with high variation in performance of the network~\cite{8784466}.

Considering throughput, researchers investigated its behavior under mobile networks and studied factors potentially impairing the performance. 
Deng~\emph{et al.}~\cite{Deng2014wifi} investigate the characterization of multihomed systems considering WiFi vs. LTE in a controlled experiment. 
They show that LTE can provide higher performance than WiFi, also exhibiting large variability on both short and long time scales.
Rathnayake~\emph{et al.}~\cite{rathnayake2012emune} demonstrate how a prediction engine can be capable of forecasting future network and bandwidth availability, and propose a utility-based scheduling algorithm which uses the predicted throughput to schedule the data transfer over multiple interfaces from fixed nodes.
Safari~\emph{et al.}~\cite{KHATOUNI201876} demonstrate the higher performance variation for moving user traveling in the vehicle via measurement. They opted to use a custom heuristic for video upload application. 
Midoglu~\emph{et al.}~\cite{midoglu2018monroe} show that the parameters in speed test application configurations, e.g., the number of TCP flows, measurement duration, and server location on measured downlink data rate can significantly affect mobile network measurement results.
Safari~\emph{et al.}~\cite{khatouni2017speedtest} illustrates the high variation in mobile network performance through a large-scale measurement. They show the performance and network setting in the European mobile networks is highly dynamic and hard to predict. 
Nunes~\emph{et al.}~\cite{nunes2011machine} focus on TCP performance in a Mobile Ad Hoc network and propose a machine learning technique called \textit{Experts Framework}. 

Finally, it is vital to understand the users' QoE.
Asrese~\emph{et al.}~\cite{Web-MBB-QoE} present a measurement tool to measure web latency and QoE in the cellular network. 
They show that the DNS lookup time and Page Load Time (PLT) of the visited websites have similar performance in LTE and fixed-line networks. 
However, the TCP three-way handshake time and Time To First Byte (TTFB) of the considered websites are longer in LTE networks.
However, the authors in~\cite{Monroe-QoE} illustrate the high randomness in mobile network performance dominates over the benefit of some new protocols, e.g., HTTP2 over HTTP1.1.
Casas~\emph{et al.}~\cite{youtube-QoE} study the problem of how to extract YouTube performance indicators related to the QoE perceived by end-users and they show the potential and feasibility of doing real-time QoE monitoring in services such as YouTube in mobile networks.
Cecchet~\emph{et al.}~\cite{mBenchLab} present an open-source infrastructure to measure the QoE of Web application on mobile devices. They illustrate the mobile network performance is a dominant factor in user-perceived QoE over device performance or user location. 
Hosek~\emph{et al.}~\cite{MobilewebQoE} illustrate QoE for mobile web service in different network conditions and highlight the impact of network parameters on mobile web QoE.
Balachandran~\emph{et al.}~\cite{balachandran2014modeling} show the impact of low-level measurable radio network characteristics on the user QoE during web browsing. 
The WebLAR tool measures web QoS and web QoE and it shows the TCP handshake time and Time To First Byte mobile networks longer than fixed-line networks using measurement in different mobile networks.
Moreover, it demonstrates QoE across several operators~\cite{andra_PAM}. In~\cite{Network-Layer-QoE}, authors model the QoE of YouTube Live service from packet-level data in a Wi-Fi network. 
Similarly, Oliver-Balsalobre~\emph{et al.}~\cite{OliverBalsalobre2017AST} model key service performance indicators from TCP/IP metrics for video-streaming (YouTube) in a Wi-Fi network has been presented. 
These are different from this work in several aspects, they focus on modeling the video stream QoE in Wi-Fi networks. Importantly, they emulate the wireless network by using a limited combination of performance metrics namely, packet loss, packet delay, and maximum throughput.  

In this context, network emulation plays a crucial role in network testing, and it is a well-studied topic in the literature. The Linux operating system has been including \texttt{tc-netem}, a tool for this purpose for more than 20 years. 
Several testing applications rely on it to install shaping policies reflecting different network conditions. 
Commercial products~\cite{netemul1,netemul2} emerged to offer software and hardware solutions for network emulation and testing, based on profiles decided at design-time defined by several parameters. Differently from them, we offer \TOOL as a freeware solution, exposing high-level network models describing the intrinsic variability of the network.  

In Table~\ref{tab:available_tools}, we summarize the mobile network profiles offered by six freeware testing platforms. We include tools used to test web pages (Google Chrome, WebPageTest, BrowserTime), operating systems (Android Emulator) and general-purpose frameworks (Advanced Traffic Control and Network Link Conditioner). 
They provide profiles decided at design-time, that simply impose simple traffic shaping policies, with limited variability. 
Moreover, they provide values highly heterogeneous, with, for example, 3G having download bandwidth from 750 kbits to 14 Mbits depending on the tool. 
In our work, we provide shaping values based on real-world measurements and offer a tool able to emulate not only plausible values but also the observed variability. 

Several research works try to build realistic emulators of mobile networks. Many works aimed at achieving realism with fine-grained models of network devices. For instance, NIST Net~\cite{carson2003nist} is a flexible and powerful emulator for WANs, while authors of~\cite{ivanic2009mobile} exploit advanced modeling methods to build a mobile network emulation environment. 
Beuran~\emph{et al.}~\cite{QOMET} propose a multi-purpose wireless network emulator to emulated 802.11 a/b/g WLAN technology over wired networks for several types of scenarios. Veltri~\emph{et al.}~\cite{Nemo} design an emulator to emulate a general IP network either a single link or a portion of a network. 

More recently, mobile network emulation followed the general trends of research in networking, and has been used to mimic real deployments of e.g., Software Defined Networks~\cite{fontes2015mininet} or flying vehicles networks~\cite{rosati2015dynamic}. 
There are few research studies building on a data-driven approach to obtain a realistic emulation. Noble~\emph{et al.}~\cite{noble1997trace} propose a pure trace-driven network emulation that re-creates the observed end-to-end characteristics of a real wireless network. KauNet~\cite{garcia2007kaunet} provides pattern-based emulation with a higher level of detail, controlling the behavior of each individual packet. Rupp~\emph{et al.}~\cite{book-Rupp} present the Vienna Simulator Suite mobile network simulators at access link level in different levels and scenario. In contrast to previous works, we build an emulator based on a large measurement dataset of operational mobile networks and make it available to the community and compatible with the \texttt{tc-netem} toolset.
\section{Measurements setup}
\label{sec:measurement}
In this section, we briefly describe the \monroe platform that we use to gather the measurements as well as the collected datasets. We also provide guidelines and source code to run a new measurement campaign. 
Table~\ref{tab:datasets} briefly describes the datasets we use. 

\begin{table}[h]
    \centering
    \footnotesize
    \caption{The high-level description of the employed datasets.}
    
    \begin{tabular}{lp{65mm}}
        \toprule
        \bf Name    & \bf Description  \\
        \midrule
        \META   & \monroe node metadata, reporting RAT, signal strength and other physical layer conditions over time \\
        \HTTP   & HTTP downloads performed periodically by \monroe nodes using the \texttt{curl} tool, used to validate \TOOL \\
        \SPEED  & Speed tests over TCP measuring download and upload bandwidth, and latency from 20 \monroe nodes\\
        \hline
    \end{tabular}
    \label{tab:datasets}
\end{table}

\subsection{Measurement infrastructure}
\label{measurement:monroe}
\begin{figure}[t]
	\centering
	\includegraphics[width=0.7\columnwidth]{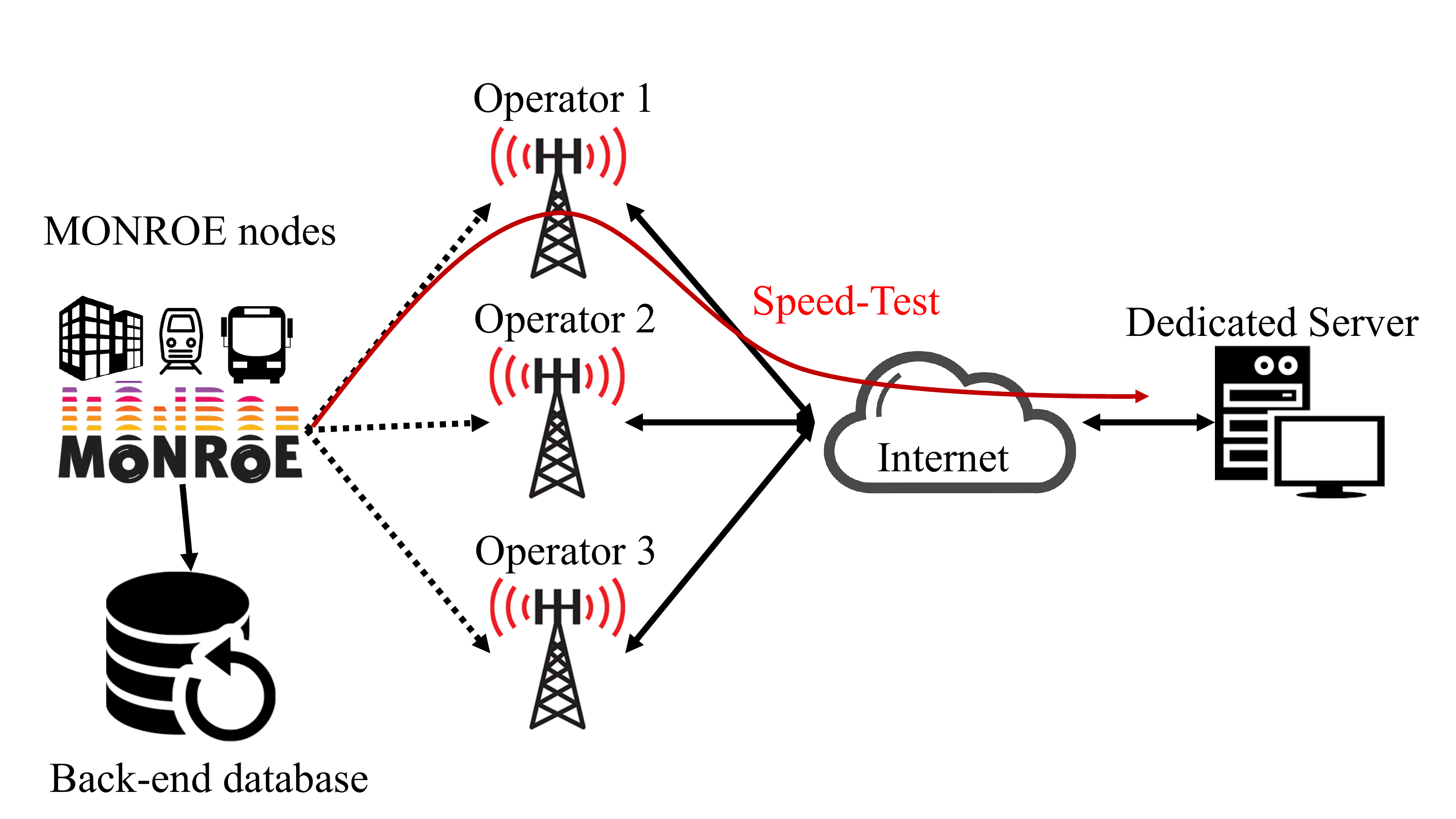}
 	\caption{The experimental setup for speed test measurement in the \monroe platform.}
	\label{fig:experiment_setup}
\end{figure}

Systematic, repeatable measurements are crucial for evaluating network performance and assessing the quality experienced by end-users. As such, researchers have built several platforms dedicated to broadband networks, e.g., RIPE Atlas\footnote{\url{https://atlas.ripe.net/}}, CAIDA Ark\footnote{\url{http://www.caida.org/projects/ark/}}, or PlanetLab\footnote{\url{https://www.planet-lab.org/}}. 
In contrast to them, \monroe~\cite{alay2017mobicom,MANCUSO201989} is a unique platform that enables controlled experimentation in a mobile network environment with different commercial mobile operators. 
In more detail, each \monroe node is equipped with different 3G/4G mobile broadband modems each one connecting to a different operator. Figure~\ref{fig:experiment_setup} reports a schematic view of a specific scenario on \monroe platform. 
Each node allows researchers to run custom experiments, to schedule measurement campaigns, and collecting data from operational mobile networks and Wi-Fi networks together along with full context information. 
The \monroe infrastructure covers $4$ countries in Europe (Italy, Norway, Spain, and Sweden) with more than $100$ nodes/locations. 

The \monroe platform allows us to access the information about network, time and location of experiments, as well as metadata from the mobile modems including, e.g., signal strength, RAT, cell identifier for each network provider.\footnote{The description and detail information about metadata is available on \url{https://github.com/MONROE-PROJECT/UserManual}} 
We collect and use metadata, and call them the \META dataset. 
The \monroe Alliance\footnote{\url{https://www.monroe-project.eu/become-a-member/}} offers different plans for new members and partners to have new nodes to be installed in their countries/locations. 
Moreover, all necessary software to replicate a \monroe node and base experiments are open source.    
Each node performs a set of predefined experiments (e.g., ping, passive monitoring, etc.) continuously. 
Experiment results are collected in the central database and stored in a Hadoop-based cluster. 
In the following, we provide details about the experimental setup and describe the considered metrics.

\subsection{HTTP download measurements}

Among other predefined experiments, \monroe nodes perform simple HTTP downloads periodically. Using the popular \texttt{curl} tool, each node downloads a large-sized HTTP object from an \emph{ad hoc} server located in the same country, e.g.,, Sweden.
The object has a size of $100$ MB and is typically downloaded in a few seconds, depending on the network conditions. 
Once the download ends, the node computes statistics such as download speed and duration, as well as several TCP-level metrics. Periodically, experiment results are sent to the central database. In total, $84$ nodes/locations performed $19$k downloads for 1 year using $10$ mobile network operators. In this work, we make use of these data and call them the \HTTP dataset.

\begin{table}[h]
    \centering
    \footnotesize
    \caption{Employed speed test measurements. We use data from $2$ countries and $4$ operators.}
    \label{tab:speedtests}
    \begin{tabular}{llrrr}
        \toprule
        \bf Country & \bf  Operator &\bf  Nodes & \bf    3G &  \bf    4G \\
        \midrule
        Norway & Telenor    &  $40$ & $495$     &  $19\,707$ \\
               & Telia      &  $35$ & $769$     &  $13\,373$ \\
               & Net1       &  $28$ & -         &  $11\,800$ \\
        Sweden & Telenor    &  $47$ & $1\,156$  &  $17\,222$ \\
               & Telia      &  $59$ & $1\,969$  &  $23\,077$ \\
               & 3          &  $46$ & $3\,347$  &   $8\,837$ \\
        \bottomrule
    \end{tabular}
\end{table}

\subsection{Speed test measurements and dataset}
\label{sec:speedtest-meas-dataset}
For the speed test measurements, we use the dataset collected in~\cite{midoglu2018monroe}. The authors built a custom measurement tool based on Open-RMBT~\cite{open-rmbt} that they called \emph{MONROE-Nettest}. It measures latency, downlink, and uplink capacity using multiple TCP connections, plus it measures the distance in terms of network hops by using \texttt{traceroute}. 
We refer the reader to~\cite{midoglu2018monroe} for further details about \emph{MONROE-Nettest} design. 
The authors used \emph{MONROE-Nettest} to run a large-scale measurement campaign involving $4$ operators on $2$ countries, namely Sweden and Norway.
The experiments were conducted from $20$ \monroe nodes/locations towards $2$ servers \emph{ad hoc} deployed in the respective countries. 
The \texttt{traceroute} measurements collected along each experiment allow us to infer the distance between the node and the server, hence the complexity of the traversed network. Results show that the traffic from the mobile node to the server never leaves the monitored country. Moreover \texttt{traceroute} measurements show that typically packets transit via $5-7$ hops.
Over one year, they performed more than $100$ k speed tests, including different hours of the day and days of the week. We provide a breakdown of the dataset in Table~\ref{tab:speedtests}. Relevant to our analysis, the measurements were taken under diverse network conditions, with both 3G and 4G RAT and different levels of signal quality. We call this dataset \SPEED.

\subsection{New dataset creation}
\label{sec:data-acquisition}

Here, we briefly summarize how other researchers can run new measurement campaigns to gather new datasets, leaving the possibility to contribute to the \TOOL project. To this end, researchers need a dedicated server and a mobile node attached to a Mobile Network Operator. The mobile node is used (i) to perform speed test measurements toward the dedicated server, and (ii) to gather network-level metadata i.e., the RAT in use and the signal strength.

For steps (i) and (ii), we recommend the \monroe platform\footnote{The \monroe codes to replicate the platform is available at \url{https://github.com/MONROE-PROJECT}
} offering the \emph{MONROE-Nettest} tool~\cite{midoglu2018monroe} to gather speed test measurements and exposing network-level metadata about each speed test~\footnote{The description is available at~\url{https://github.com/MONROE-PROJECT/data-exporter}}. However, \TOOL can be built on top of any automatic framework for mobile network measurements. Hence, an user willing to run new measurements, could either rely on this code repository or perform speed tests independently. 

The results of the experiment campaign must produce a \texttt{csv} file describing the speed tests along with metadata information. To check the format of the file, in the \TOOL repository~\cite{mne} we report an example of it. The repository also contains the code to create new \TOOL models from the \texttt{csv} file, as we explain in the following section.
\section{\TOOL: \TOOLNAME}
\label{sec:methodology}
\begin{figure}[t]
	\centering
	\includegraphics[width=\columnwidth]{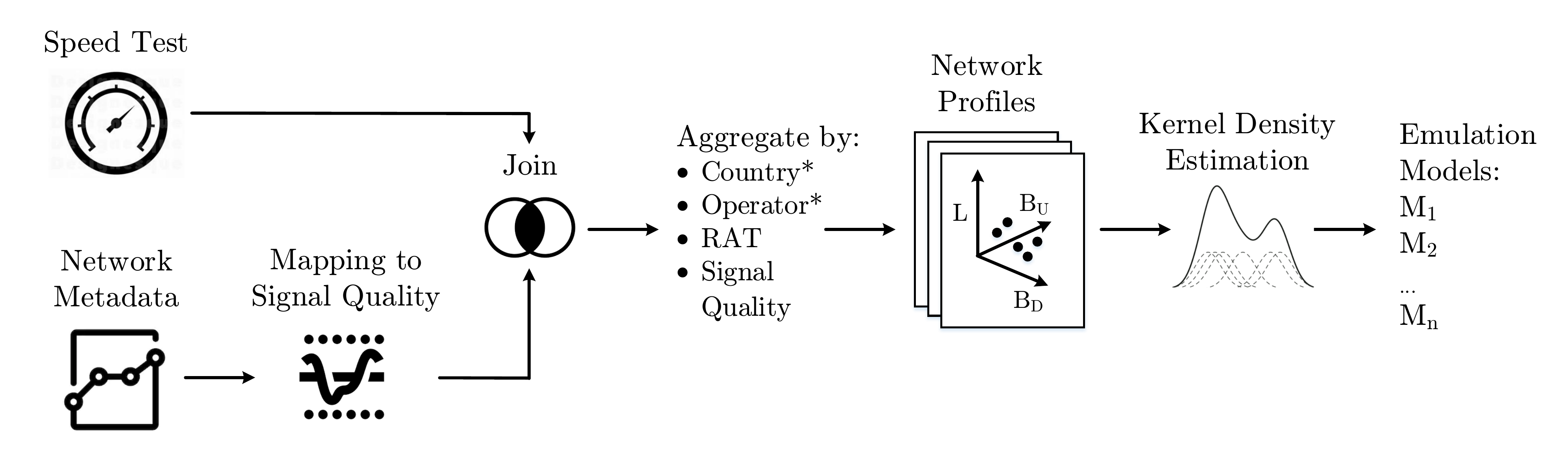}
 	\caption{Generation of \TOOL models. Fields with * are used only for specific profiles and not to build universal ones.}
	\label{fig:tool_training}
\end{figure}

In this section, we illustrate the design of \TOOL. First, we describe all the data processing steps required to clean and join the \META and \SPEED datasets to create network profiles. Then, we describe how we build our network emulation models based on the available profiles. Finally, we explain how a tester can use \TOOL models to achieve realistic traffic emulation. Figure~\ref{fig:tool_training} provides an overview of the model creation process.

\subsection{Data cleaning and processing}

Armed with the \META dataset, we first map the signal strength values in three \textit{signal quality} levels: bad, medium, and good. Table~\ref{tab:bins} reports the binning boundaries.\footnote{Boundaries are defined based on \url{https://wiki.teltonika.lt/view/Mobile_Signal_Strength_Recommendations.}} 
Then, we join the \SPEED and the \META datasets to augment each speed test measure with contextual information (i.e., country, operator, RAT, signal quality). Given the large size of the datasets, most of the processing here is performed using an Apache Spark cluster. 
Here, we discard a small number of speed tests due to the lack of physical layer information caused by temporary issues of the nodes. At this point, we aggregate measurements taken under the same conditions to build the network \textit{profiles}. In particular, we build \textit{specific} network profiles for country and operator by aggregating the \SPEED data by:
\begin{itemize}
    \item \textit{Country:} where the measurements were taken
    \item \textit{Operator:} which operator was used 
    \item \textit{Radio Access Technology:} 3G or 4G
    \item \textit{Signal Quality:} the mapped signal strength i.e., the \textit{rssi}.
\end{itemize}

Furthermore, we create \textit{universal} profiles, containing the performance of \emph{any} operator using different RAT and signal quality, by aggregating the \SPEED data only by: \textit{Radio Access Technology} and \textit{Signal Quality}.

\begin{table}[h]
    \centering
    \footnotesize
    \caption{Binning boundaries for 3G and 4G. Values are expressed in $dB$.}
    \label{tab:bins}
    \begin{tabular}{lrr}
        \toprule
        \bf Quality    &\bf   3G   & \bf 4G  \\
        \midrule
        Bad       & $rssi \leq -100$ & $rssi \leq -85$ \\
        Ordinary   & $-100 < rssi \leq -85$ & $-85 < rssi \leq -75$ \\
        Good       & $rssi > -85$ & $rssi > -75$ \\
        \bottomrule
    \end{tabular}
\end{table}

After discarding all profiles with less than $100$ measurements i.e., those not having enough samples to be reliable, we end up with $32$ network profiles: $26$ for specific operators and countries and $6$ universals. They include data from $2$ countries and $4$ distinct operators ($2$ of which are present in both countries). 
They also differ for the signal qualities, in particular we have $6$ profiles with \emph{bad} quality, $13$ with \emph{medium} quality, and $13$ with \emph{good} quality. Considering RAT, we have $14$ profiles for 3G and $18$ for 4G. As physical conditions (signal quality and RAT) could not be controlled when measurements were taken, we have a significant imbalance in the quantity of per-profile samples. 

\begin{figure*}[h!]
    \begin{center}
        \begin{subfigure}{0.49\textwidth}
            \includegraphics[width=\columnwidth]{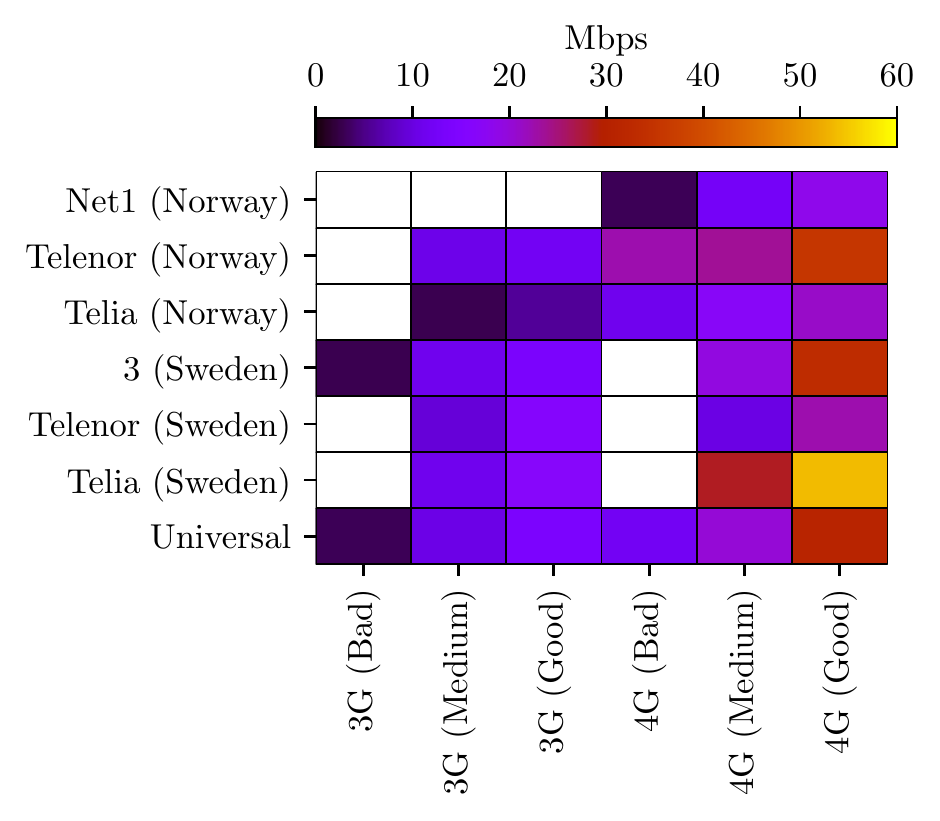}
            \caption{Download.}
            \label{fig:profiles_down}
        \end{subfigure}
        \begin{subfigure}{0.36\textwidth}
            \includegraphics[width=\columnwidth]{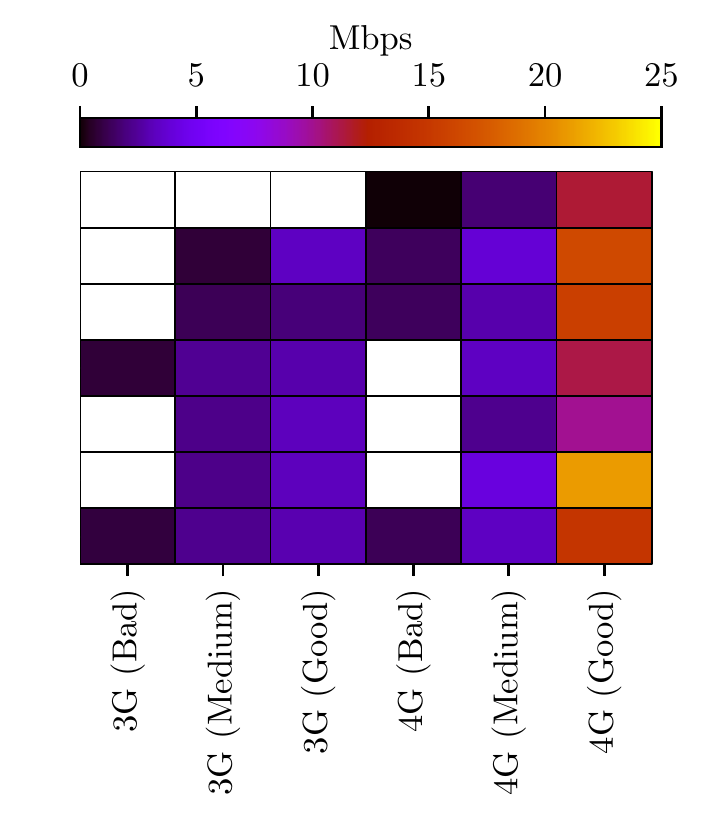}
            \caption{Upload.}
            \label{fig:profiles_up}
        \end{subfigure}
        \begin{subfigure}{0.49\textwidth}
            \includegraphics[width=\columnwidth]{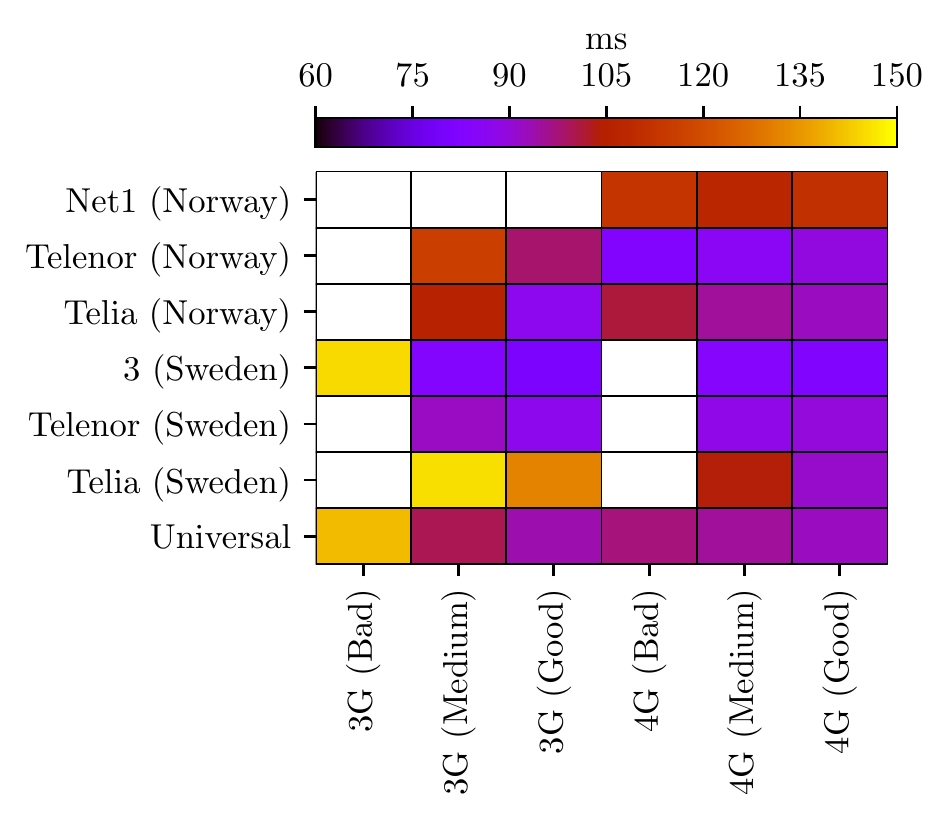}
            \caption{Latency.}
            \label{fig:profiles_latency}
        \end{subfigure}
        \caption{Overview of the obtained profiles. Only median values are reported, while variability is omitted.}
        \label{fig:profiles}
    \end{center}
\end{figure*}

To illustrate the available network profiles and show their difference, in Figure~\ref{fig:profiles}, we report for each one the median value of download bandwidth, upload bandwidth, and latency. 
Focusing on the download and upload bandwidth, we notice how median values are not sufficient to strongly differentiate different 3G profiles, while good 4G profiles strongly differ from each other. On the one hand, this highlights the importance of having dynamic profiles, as static values would not well differentiate between different 3G networks. 
On the other hand, despite the universal profiles give a general idea of the possible performance, it is also important to define specific profiles to capture the differences among operators or countries. 

It is interesting to notice that the measured values are in all cases far from the nominal data rates of the radio standards. For 3G, download throughput seldom exceeds 20 Mbit/s while the HSPA+ standard should reach up to 42.2 Mbit/s~\cite{korowajczuk2011lte}. The difference is even larger for 4G, where the LTE specification provides downlink peak rates of 300 Mbit/s and uplink peak rates of 75 Mbit/s~\cite{korowajczuk2011lte}, while we observe values in the order of $20-60$ Mbit/s.

\begin{figure*}[t]
    \begin{center}
        \begin{subfigure}{0.45\textwidth}
            \includegraphics[width=\columnwidth]{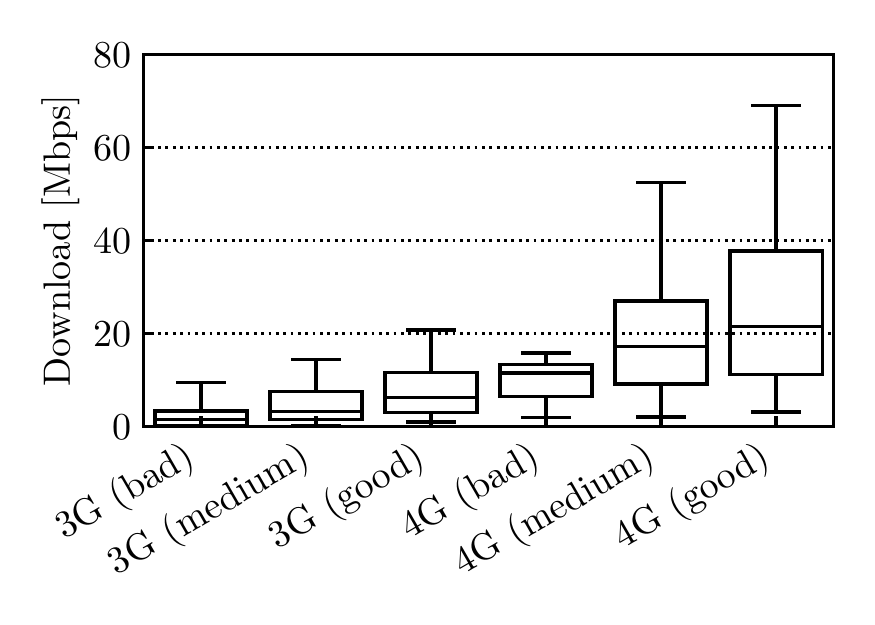}
            \caption{Download.}
            \label{fig:cdfs_1}
        \end{subfigure}
        \begin{subfigure}{0.45\textwidth}
            \includegraphics[width=\columnwidth]{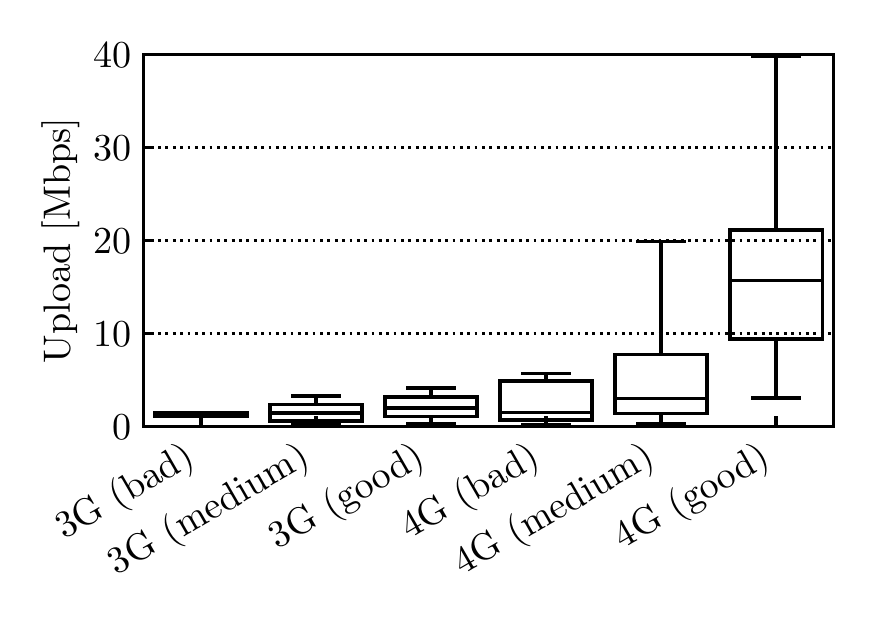}
            \caption{Upload.}
            \label{fig:cdfs_2}
        \end{subfigure}
        \begin{subfigure}{0.45\textwidth}
            \includegraphics[width=\columnwidth]{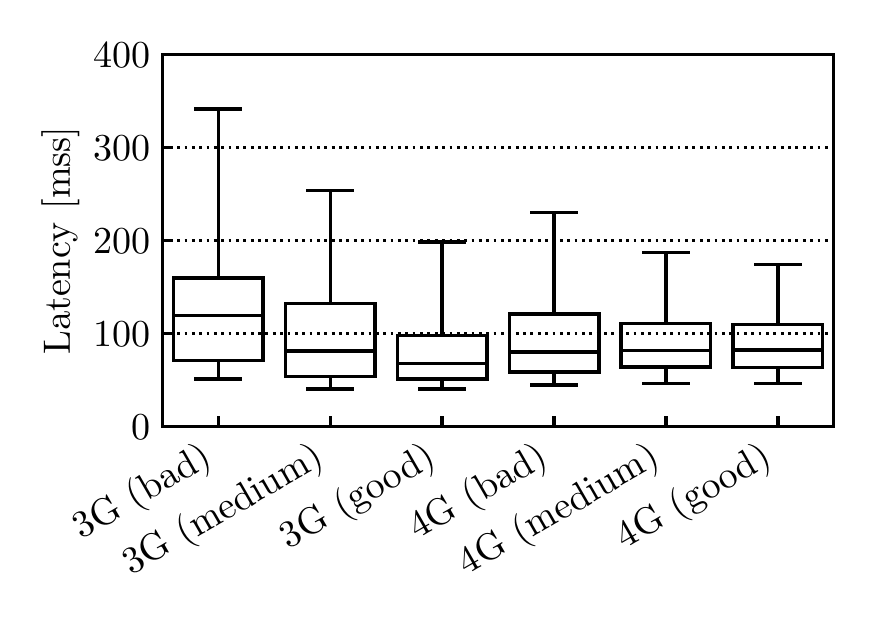}
            \caption{Latency.}
            \label{fig:cdfs_3}
        \end{subfigure}
        \caption{Speed test results for Telia (Norway).}
        \label{fig:cdfs}
    \end{center}
\end{figure*}

To underline the importance of variability of network conditions, in
Figure~\ref{fig:cdfs}, we show the distribution of the measured values for download bandwidth (Figure~\ref{fig:cdfs_1}), upload bandwidth (Figure~\ref{fig:cdfs_2}) and latency (Figure~\ref{fig:cdfs_3}) for the Telia operator in Norway. 
We report separately profiles obtained with different combinations of RAT and signal quality. Boxes in the figure span from the first to the third quartile, whiskers from the 5\textsuperscript{th} to the 95\textsuperscript{th} percentile and the black strokes represent the median values. 
The impact of physical layer conditions appears clear. Considering download bandwidth in Figure~\ref{fig:cdfs_1}, 4G performs better than 3G, e.g., good 4G offers in median 20 Mbits in download, while good 3G only 6 Mbits. 
The role of signal quality is sizeable too, with good 4G almost twice as fast as bad 4G. Variability is high in all cases, mandating a realistic emulator to consider it. 
For latency (Figure~\ref{fig:cdfs_3}), the impact of physical layer conditions is limited, with similar median values, except for bad 3G. Again, variability is rather high, with the inter-quartile range in the order of tens of milliseconds. Similar considerations hold for other profiles, not reported here for the sake of brevity. However, Figure~\ref{fig:profiles} still shows the overall absolute values, while the variability for all profiles can be found in the \TOOL repository~\cite{mne}. 
This result emphasizes the high variability in mobile networks and high dependency on location, operator, technology, and signal quality. Our data represent a large set of operators in more than 80 locations with two RATs but clearly do not include all possible scenarios. As such, the open-source \monroe platform and \TOOL gives the opportunity for users to reproduce the same methodology and realistically simulate their local operators, other than sharing their datasets with the community. 5G technology is testing and installing around the world and it calls for performance analysis and simulators. For this, the \monroe Alliance~\cite{Monroe-5G} is working to equip and adapt the platform to 5G technology, giving the possibility to have new 5G profiles in \TOOL in the future. 

\subsection{Emulation model creation}

Given the different network profiles, here we use the KDE methodology~\cite{parzen1962estimation} to build our emulation models, each one modeling a different profile. From a profile $p$, we extract the network parameters measured by each speed test in the form of download bandwidth ($B_D$), upload bandwidth ($B_U)$, and latency ($L$). Although the characteristics of a network might be more complex to model (e.g., due to jitter, queue size, etc.), the speed test measures only provide these three parameters. Moreover, these are more straightforward to tune and several network emulation tools use only these three values to define a network behaviour.
As a result each speed test measurement in $p$ is described by a tuple in the form: $(B_D,B_U,L)$. Instead of merely sampling from the observed $(B_D,B_U,L)$ values, we build our dynamic \textit{emulation model} $M_p$ using the KDE methodology.

KDE is largely used in the literature as a tool to model multivariate problems~\cite{GERBER2014115,kde_road_patterns}. In this work, we use the KDE to estimate the multivariate probability of observing $(B_D,B_U,L)$ in a speed test measurement for profile $p$. We fit a three-dimensional KDE on all possible value of $(B_D,B_U,L)$ for each profile. In this way, we obtain distinct models describing the network parameters for each profile. We use a Gaussian kernel to define the influence of each speed test measure and set the bandwidth with the Silverman method~\cite{Silverman86}. For the implementation, we rely on the SciPy Python module. In a nutshell, we use KDE as a data modeling tool able to capture performance patterns from speed tests, hence giving us the possibility to abstract performance samples while reducing the impact of noise or secondary finer-scale phenomena. 

This said \TOOL considers all the measurements collected for profile $p$, and estimates their multivariate probability function to build $M_p$. The novelty of \TOOL is twofold. First, the emulation models are based on a large-scale measurement campaign, including $100$ k speed test measurements. Second, \TOOL exposes models and not static values. Hence \TOOL can correctly replicate the variability of the network conditions as observed in the real \SPEED dataset.

We provide the code to create new models given a set of speed test measurements in the repository~\cite{mne}. This code, written in Python, takes as input a \texttt{csv} file with the speed test measurements along with their metadata information as described in Section~\ref{sec:data-acquisition}. As a result, it creates an output file containing the models directly usable by \TOOL. Using this piece of code the user can translate an experimental campaign into additional models describing different operators, different locations, mobility, different RAT like the 5G access technology.

\subsection{Emulation model usage}

We now describe how \TOOL uses the models at run time, to dynamically emulate different network conditions. At run time \TOOL offers three possibilities to sample new values for the tuple $(B_D,B_U,L)$ from model $M_p$. 

Firstly, at each run \TOOL can extract different values for the three key parameters. In other words, at each run, \TOOL samples potentially different parameters, following the observed distribution. As such, the more likely certain values appear in real data, the more frequent will be sampled from the distribution. The main goal of \TOOL is to allow the user to run \emph{several} experiments using a given model $p$, and, consequently, to observe the \emph{distribution} of the obtained effects. We will use this solution to validate our models in Section~\ref{sec:model_validation}.

Secondly, \TOOL allows the tester to sample new emulation values every $n$ seconds to emulate dynamic conditions within an experiment e.g., testing watching video while moving. Note that we cannot provide data-driven guidelines on how to set $n$ as we have not performed experiments using mobile nodes, and, as such, we leave the choice of this parameter to the user. However, studying mobile networks in their dynamic behavior during mobility is of great interest, but we leave this for future work. A typical use case is testing the performance of adaptive protocols such as the Dynamic Adaptive Streaming over HTTP under different mobile network scenarios. We explore this and other use cases in Section~\ref{sec:web_use_case}.

Finally, \TOOL also offers the possibility to change network profiles within an experiment following a \emph{trace-driven scenario} written in a configuration file. By using the convenience script \texttt{trace\_run}, the tester can run an experiment following a scenario containing potentially drastic networking changes e.g., a user going through a tunnel, or moving from an area with good coverage to a bad one. This feature can be used to evaluate the application performance in stressful stations such as a connection migrating from a Good 4G quality, down to a Bad 3G one, and back to a 4G Medium situation. 

Under the hood, \TOOL uses the \texttt{tc-netem} Linux tool to impose traffic shaping policies on the selected network interface. In particular, we impose bandwidth limitations by means of the Hierarchy Token Bucket (HTB) a classful \textit{qdisc}, while we use the \texttt{netem} \textit{qdisc} to emulate latency. We make use of the Intermediate Functional Block (\texttt{ifb}) kernel module to create the pseudo network interface necessary to control the incoming bandwidth. We validate \TOOL accuracy in emulating network conditions in Section~\ref{sec:emulation_resuts}.

We release \TOOL as open-source, and it is available online~\cite{mne}. It consists of a simple tool that installs traffic shaping policies according to the model requested by the user in one of the above solutions on the desired network interface. 
\section{Experimental results}
\label{sec:emulation_resuts}

In this section, we analyze the performance of \TOOL, comparing the results we obtain with those collected in a real environment and with other freeware emulation tools. We then study the impact of the number of speed test measurements on the created models to suggest the minimum number of experiments required to create a model. 

\paragraph*{\SIMPLE Baseline}

In the following, we present \TOOL results along with those obtained with a simpler emulation method, that we use as a baseline. We call it the \SIMPLE, and it consists of a basic script that uses \texttt{tc-netem} to install the desired shaping policies. While \TOOL samples new shaping parameters at each run, our baseline approach only uses the average observed values of download and upload bandwidth for the given model. To recreate variability, we impose a Gaussian latency distribution with the average and standard deviation values observed in the empirical distribution. Note that \texttt{tc-netem} allows variability only in latency, while there is no straightforward way of varying the bandwidth automatically. We show that a simple approach based on the Off-The-Shelf features of \texttt{tc-netem} fails to provide realistic results when it comes to real applications such as web browsing or video streaming.

\subsection{Model validation}
\label{sec:model_validation}

In this section, we validate our emulation models showing that with \TOOL a tester can accurately emulate mobile network conditions, leading to results comparable with those obtained in real networks. Here, we do not provide results about the accuracy of \TOOL in installing accurate traffic shaping rules, as for that we rely on the state-of-art tool \texttt{tc-netem}. Rather, we show that running a simulation campaign using \TOOL allows the user to observe the same distribution of desired effects that would observe in a real mobile environment.

We rely on the \HTTP dataset to validate \TOOL accuracy on replicating the network variability. The \HTTP dataset allows us to test the models with data that were never used to build the model. We aim at understanding to what extent \TOOL is able to recreate the variability of conditions experienced in real networks, in addition to the \emph{average behavior}. 
To this end, we join the \HTTP and the \META dataset, to enrich the former with the contextual information in which the HTTP downloads were performed (operator, signal quality, etc.). This allows us to map each experiment to the corresponding profile that we can emulate with \TOOL. Then, we use \TOOL to replicate the experiments emulating the corresponding network conditions -- i.e., to download large-sized HTTP objects. This is done for all profiles for which we find real data in the \HTTP dataset, $12$ in total. 

Then, for each profile, we download $1\,000$ times a 10 MB HTTP object from a dedicated server. At each download, we run \TOOL to sample new parameters of $(B_D,B_U,L)$ for the corresponding model. These experiments aim at evaluating the impact of the variability introduced by \TOOL in the final results. Indeed, within each experiment, it imposes fixed emulation parameters, but they vary across different downloads. For reference, we also use the \SIMPLE to provide a baseline.

\begin{figure*}[t]
    \begin{center}
        \begin{subfigure}{0.45\textwidth}
            \includegraphics[width=\columnwidth]{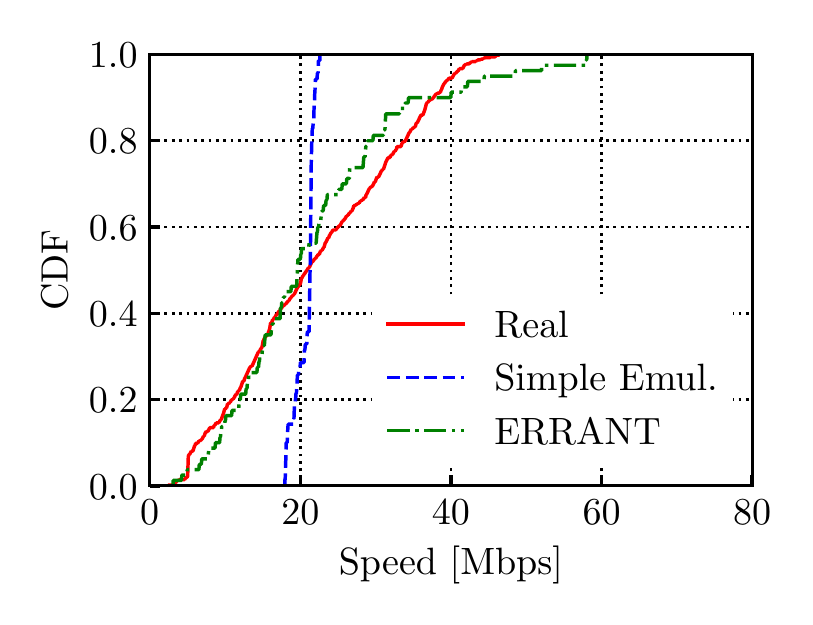}
            \caption{Telia (Sweden), ordinary 4G.}
            \label{fig:curl_cdfs_1}
        \end{subfigure}
        \begin{subfigure}{0.45\textwidth}
            \includegraphics[width=\columnwidth]{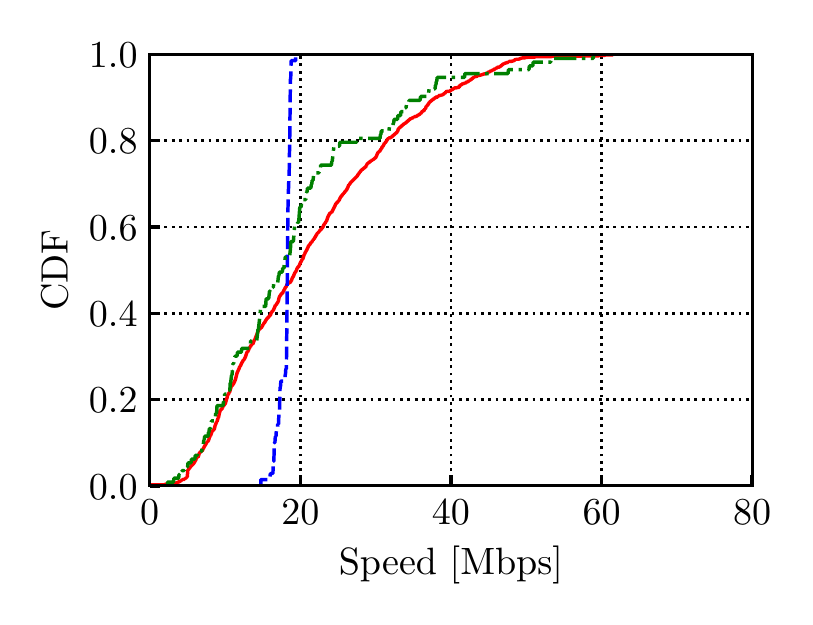}
            \caption{Telenor (Sweden), good 4G.}
            \label{fig:curl_cdfs_2}
        \end{subfigure}
        \begin{subfigure}{0.45\textwidth}
            \includegraphics[width=\columnwidth]{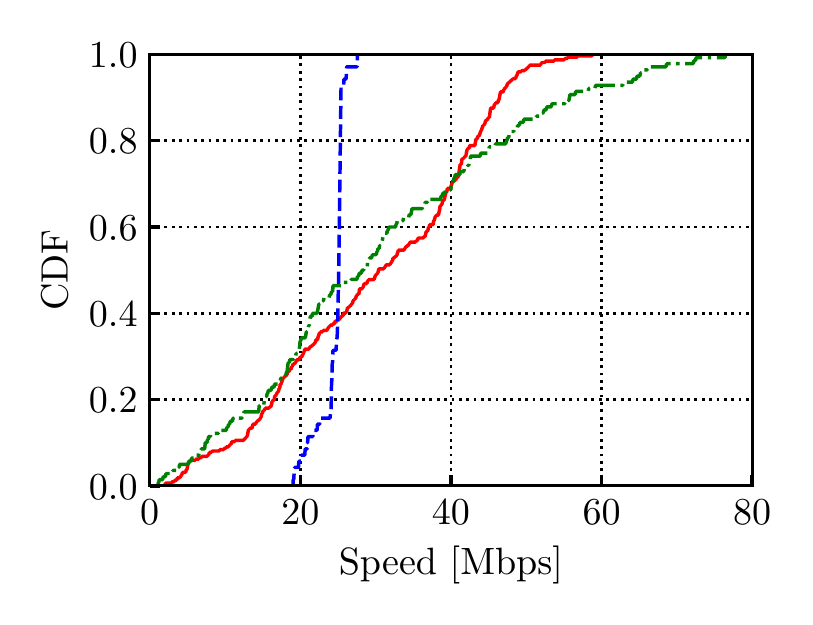}
            \caption{3 (Sweden), good 4G.}
            \label{fig:curl_cdfs_3}
        \end{subfigure}
        \caption{Comparison of HTTP download speed for 3 scenarios.}
        \label{fig:curl_cdfs}
    \end{center}
\end{figure*}

Figure~\ref{fig:curl_cdfs} presents results for three models. 
Similar results are obtained in other cases. Each plot reports the Cumulative Distribution Function (CDF) of the average download speed for the HTTP object at each run, as measured by the \texttt{curl} tool. 
We separately report results measured in real mobile networks (solid red line), using \TOOL (green dashed line), or with \SIMPLE (blue dashed line). For all profiles, the speed observed in the real case has a considerable variability, with the inter-quartile range spanning tens of Mbits. Looking at obtained results when using \TOOL, they appear very similar to the real case, considering both median values as well as the shape of the distribution. The result is different when imposing fixed network conditions with the \SIMPLE, where negligible variability is observed and where only the median value correctly describes the real data.

In conclusion, we show that \TOOL accurately recreates mobile network conditions, leading to similar results in terms of average behavior and observed variability when downloading HTTP objects. 

\subsection{Comparison with other emulation tools}

After validating \TOOL models, we compare its performance with other freeware network emulation tools already reported in Table~\ref{tab:available_tools}. 
Similar to \TOOL, these tools allow the tester to tune the download and upload bandwidth, in addition to the network latency based on different network profiles. Each tool exposes off-the-shelf network profiles, whose parameters are decided at design-time. In addition, these tools allow the user to define custom profiles based on user-specified parameters. 


Similarly to the experiments in the previous section, we use the tools reported in Table~\ref{tab:available_tools} to impose traffic shaping rules on a test machine. In parallel, we perform HTTP downloads using \texttt{curl} and record the average download speed per download. For each experiment, we download $1\,000$ times a 10 MB sized file.

\begin{figure}[t]
	\centering
	\includegraphics[width=0.7\columnwidth]{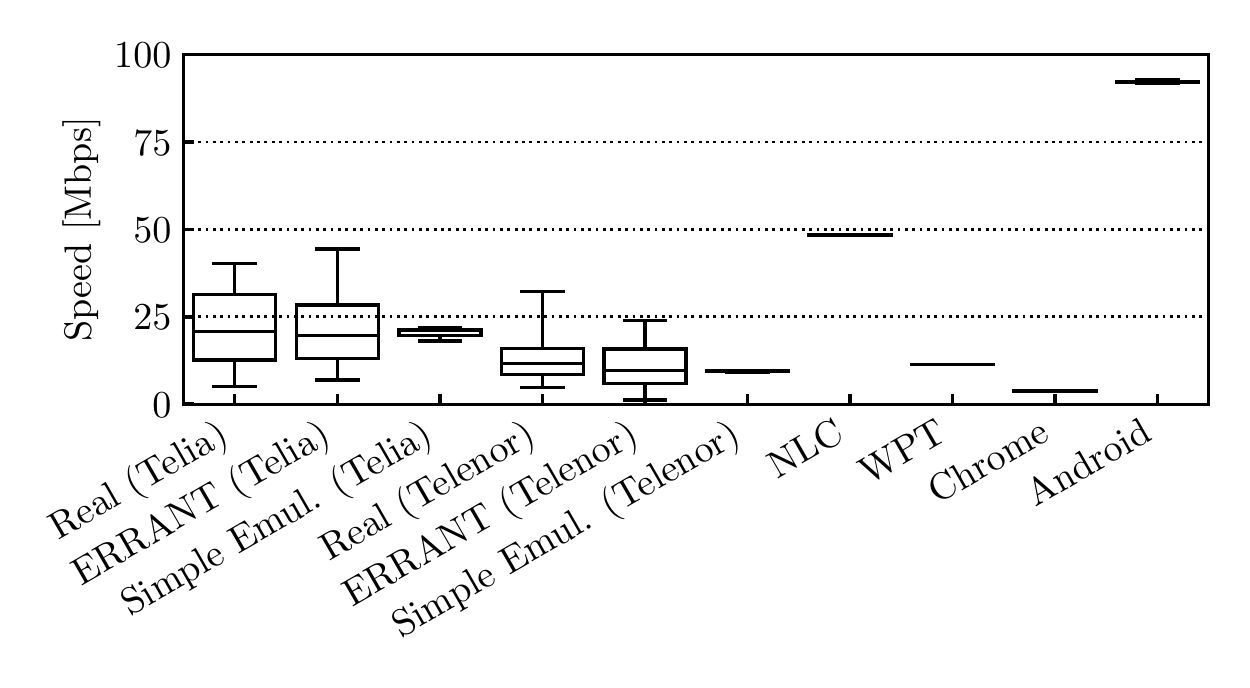}
 	\caption{Distribution of HTTP download speed in a real environment, using \TOOL and other emulation tools.}
	\label{fig:tool_comparison}
\end{figure}

In Figure~\ref{fig:tool_comparison}, we compare the performance of the 4 network emulation tools, \SIMPLE, and \TOOL with the performance observed in the real environment for two 4G (good) profiles. 
We use boxplots to represent the distribution of the download speed measured in each scenario. Boxes in the figure span from the first to the third quartile, whiskers from the 5\textsuperscript{th} to the 95\textsuperscript{th} percentile and the black strokes represent the median values. First, we note again that \TOOL leads to results similar to the real case, for both Telia and Telenor operators. The \SIMPLE instead only reproduces the average behavior but fails in providing variability in the results. Considering the four emulation tools, we first note that two of them (NLC and Android) lead to results very different from what we observe in the real networks. In particular, these two lead to a download speed of 50 and 90 Mbits respectively, while we seldom observe values higher than 30 Mbits on the real environment. WPT and Chrome instead have typical values closer with the measurements taken under the Telenor operator especially. However, we note that they provide no variability in the results, expected considering that they impose static shaping policies. 

In summary, differently from other freeware traffic shaping tools (e.g., for web page testing), \TOOL imposes network conditions based on real large-scale experiments. Allowing one to re-obtaining the variability intrinsically rooted in mobile networks. And offering models better describing the performance of real mobile networks.


\subsection{The impact of profile size}

Understanding how many speed tests are required to create a reliable model is of crucial importance to determine the size of new experimental campaigns, e.g., in the case of 5G. 
To this end, we study the impact of the number of measurements on the obtained \emph{profiles} as they are the baseline to create our models. In particular, we study to what extent a \emph{profile} composed by a small number of speed tests is representative of those obtained with the full dataset.

For this analysis, we discard profiles with less than $10$ k speed tests, that would not fit the analysis. We obtain $4$ profiles, from which we extract, by sampling, exactly $10$ k points, for a fair comparison. We now perform experiments sampling a smaller subset of samples, and compare them with the full $10$ k samples. At each run, we randomly create a subset of $n$ points per profile, where each one is described by a value of $(B_D,B_U,L)$. Then we evaluate the similarity between the subset with respect to the full set of data by means of the two-sample Kolmogorov–Smirnov test~\cite{kolmogorov1933sulla,smirnov1939estimation}, that measures the dissimilarity of two empirical distribution functions providing the $D$ statistic. For each experiment having a different subset size, we take $100$ independent subsets of $n$ points. Intuitively, $D$ measures the maximum distance between the two sets, the higher is the value the more is the dissimilarity. As such a high $D$ means that with the given subset we cannot create a profile similar with respect to the one created with the whole dataset. Hence, we cannot create a reliable model. We run our analysis separately by $B_D,B_U$, and $L$ since all proposals that adapt the Kolmogorov–Smirnov test to multivariate distributions suffer for severe scalability issues~\cite{lopes2007two}.

\begin{figure}[t]
	\centering
	\includegraphics[width=0.7\columnwidth]{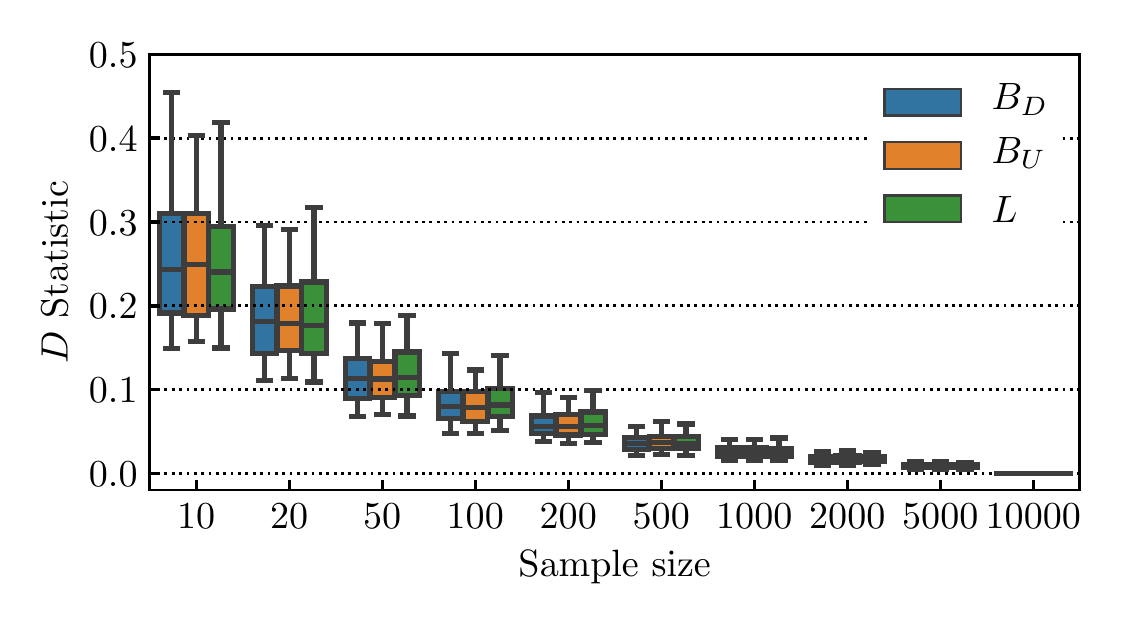}
 	\caption{Kolmogorov–Smirnov $D$ statistic when comparing the distributions obtained with $10$k with those obtained with smaller sub-samples.}
	\label{fig:ks}
\end{figure}

Figure~\ref{fig:ks} shows the distribution of the $D$ statistic when varying the sample size $n$. Clearly, when $n=10$ k, $D$ is always $0$ as the two distributions are identical. For small values of $n$ we obtain considerably different distributions, with $D$ often higher than $0.3$ when $n=10$. By increasing $n$, the distributions become more similar, with median values of $n$ reaching $0.08$ for $n=100$, and $0.02$ for $n=1000$.
These results confirm the importance of having a large dataset but show that in a new campaign with $100$ speed tests we can already have a reasonable estimation of a profile. Indeed, for $n=100$ the two sample sets come from the same distribution with a \emph{significance level} always above $0.2$, generally considered strong evidence.
As such, in \TOOL we create models only for profiles having more than $100$ points, as stated in Section~\ref{sec:methodology}. 

\section{Use cases}
\label{sec:web_use_case}

In this section, we illustrate the capability of \TOOL in providing useful insights for various stakeholders of computer networks. 
Indeed, accurate network emulation is fundamental for website designers, web application developers, and network providers to fully understand the impact of their design choices by monitoring the performance of applications before the deployment. 

\subsection{Web browsing QoE estimation}

We here focus on the use case of web browsing to show how \TOOL improves the emulation with respect to \SIMPLE.
It is crucial to understand the users' QoE especially in mobile networks which makes it more complicated as investigated by the authors of~\cite{Web-MBB-QoE,Monroe-QoE,youtube-QoE,mBenchLab,MobilewebQoE,balachandran2014modeling,andra_PAM,Network-Layer-QoE}.
Moreover, large companies operating on the web are particularly interested in studying the factors affecting QoE, with Google and Amazon reporting losses in the 0.6-1.2\% range if delay increases by 0.4-1 sec.\footnote{\url{https://www.fastcompany.com/1825005/how-one-second-could-cost-amazon-16-billion-sales}}

For our experiments, we consider three popular websites, chosen from the globally top-ranked in the Alexa ranking service, namely: Wikipedia, Google, and YouTube.\footnote{\url{https://www.alexa.com/topsites}} 
We then instrument Google Chrome to automatically visit the homepages of the selected websites using the Browsertime toolset.\footnote{\url{https://github.com/sitespeedio/browsertime}} 
In parallel, we use \TOOL to varying network conditions, using all the $32$ profiles we built. For each profile and website, we perform $100$ visits. For comparison, we also use the \SIMPLE for each profile. Finally, we visit the websites without any traffic shaping to provide a reference in the results. At each run, we collect all the metadata about the visit, including QoE-related metrics such as OnLoad time and Speed Index. Indeed these two metrics have been proved to have a high correlation to users' QoE, intrinsically subjective~\cite{da2018narrowing}. 
In total, we performed more than $18$ k visits.

\begin{figure*}[t]
    \begin{center}
        \begin{subfigure}{0.45\textwidth}
            \includegraphics[width=\columnwidth]{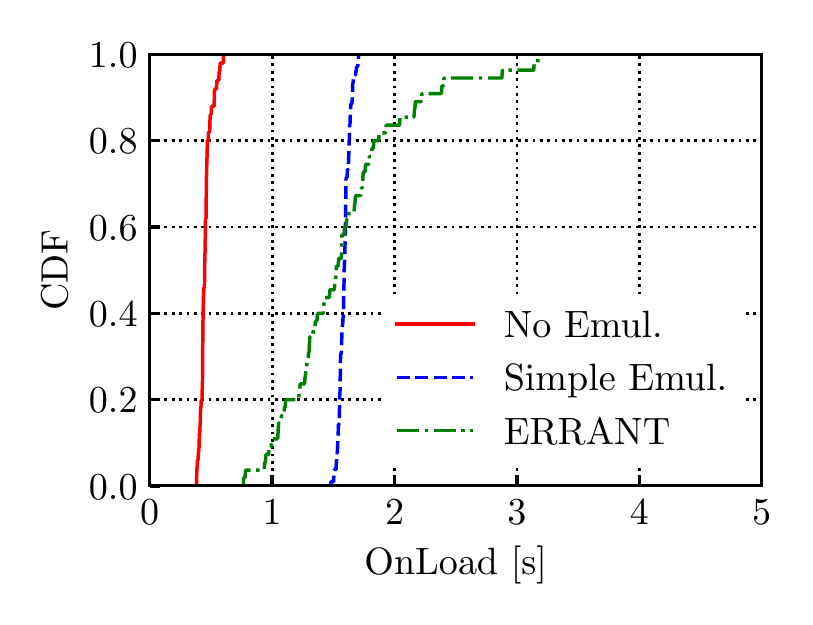}
            \caption{\emph{Wikipedia.}}
            \label{fig:web_wiki}
        \end{subfigure}
        \begin{subfigure}{0.45\textwidth}
            \includegraphics[width=\columnwidth]{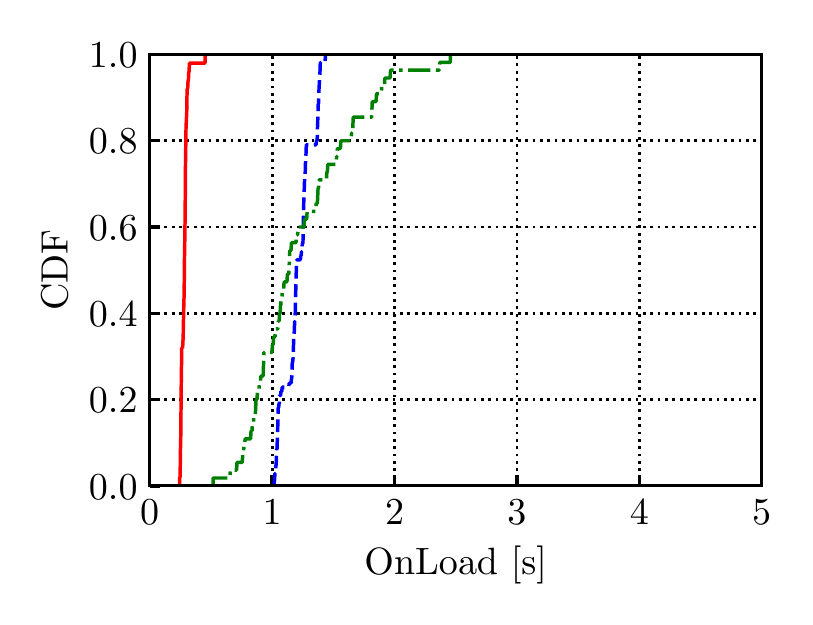}
            \caption{\emph{Google.}}
            \label{fig:web_google}
        \end{subfigure}
        \begin{subfigure}{0.45\textwidth}
            \includegraphics[width=\columnwidth]{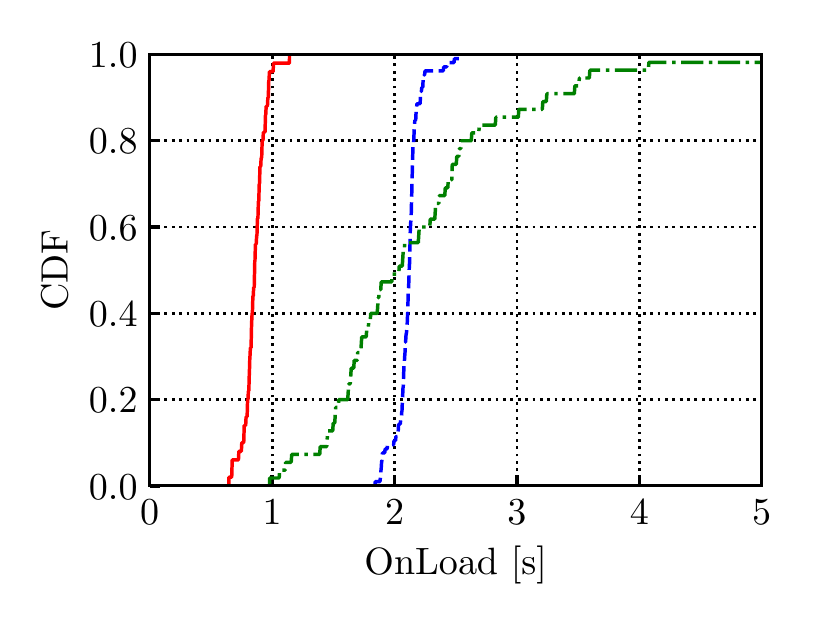}
            \caption{\emph{YouTube.}}
            \label{fig:web_yt}
        \end{subfigure}
        \caption{Comparison of \emph{OnLoad} time for 3 websites, when emulating good 4G for Telia in Sweden.}
        \label{fig:web}
    \end{center}
\end{figure*}

Figure~\ref{fig:web} shows results for \emph{good 4G, Telia (Sweden)}, while we omit the other profiles as they lead to similar conclusions. Figure~\ref{fig:web} reports the CDF of OnLoad time when using \TOOL, or static emulation or finally no emulation at all. We do not report results for SpeedIndex for the sake of brevity. 
Comparing \TOOL results (red line) with those obtained without emulation (green line), we notice how OnLoad time is strongly affected, as expected, for all websites. Instead, the \SIMPLE  (blue line) leads to values of page load time with similar median values -- e.g., $2$ seconds for YouTube. However, \TOOL imposes variability, based on the speed test data used for creating the profiles, and this variability reflects into the page load time. 
For example, YouTube page load time has an inter-quartile range of $0.8$s while only $0.1$s is observed with static emulation. Moreover, in 15\% of cases, we observe a value larger than $3$s, which makes 53\% of users to leave the website according to recent studies~\cite{an2018find}. As such, we stress the fact that \TOOL can reproduce the variability of mobile networks, allowing one to thoroughly study the behavior of applications with a level of realism previously unavailable. 

\subsection{Adaptive video streaming}

\begin{figure*}[t]
    \begin{center}
        \begin{subfigure}{0.75\textwidth}
            \includegraphics[width=\columnwidth]{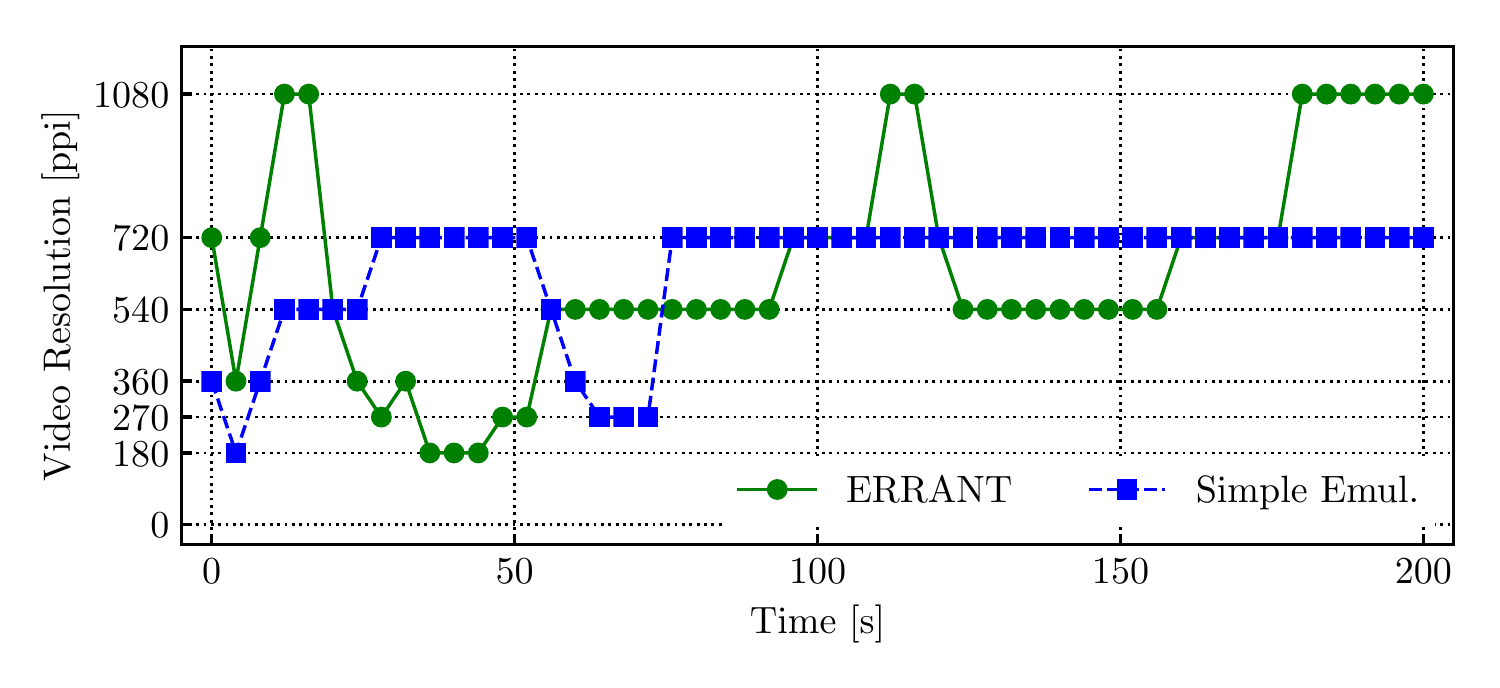}
            \caption{\emph{Bad 3G.}}
            \label{fig:video_3G}
        \end{subfigure}
        \begin{subfigure}{0.75\textwidth}
            \includegraphics[width=\columnwidth]{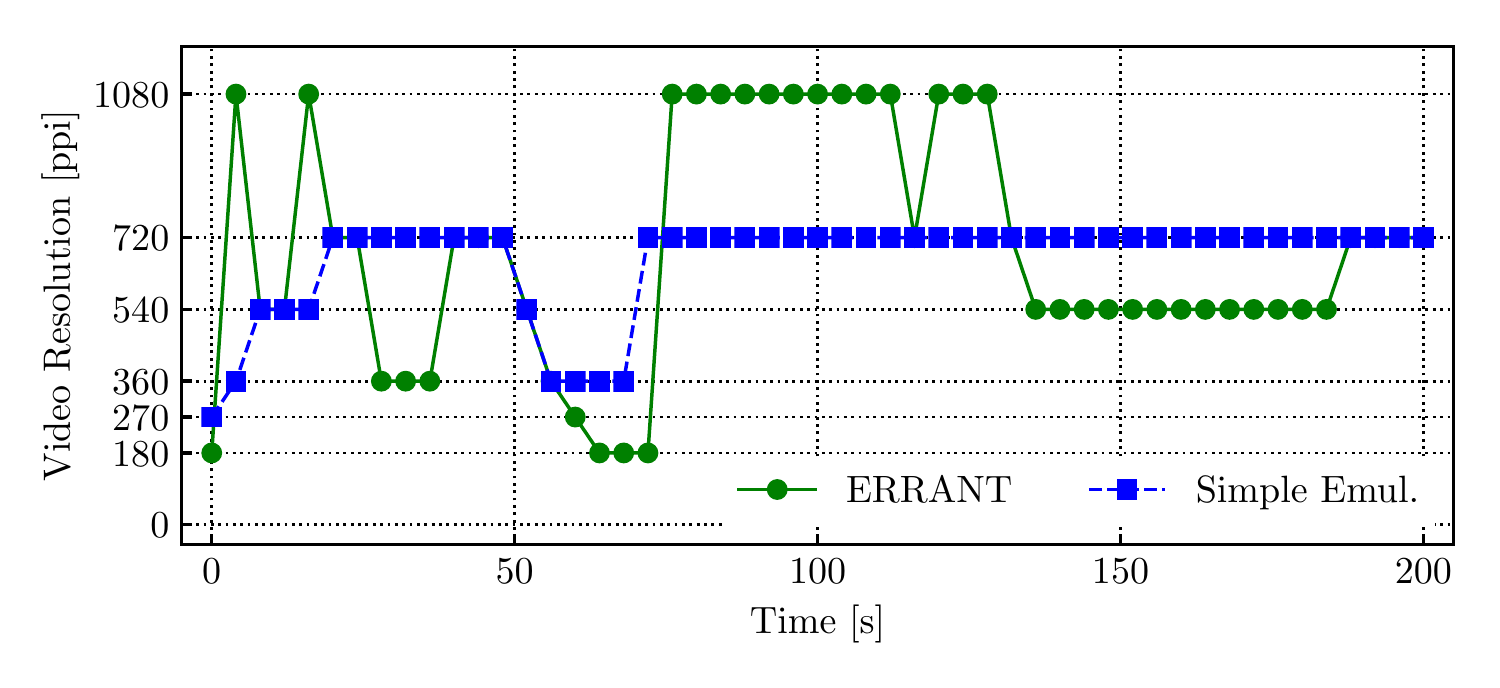}
            \caption{\emph{Bad 4G.}}
            \label{fig:video_4G}
        \end{subfigure}
        \caption{Example of adaptive streaming video sessions.}
        \label{fig:video}
    \end{center}
\end{figure*}

Here, we show a second use case, in which \TOOL is used not only to introduce \emph{inter-experiment} but also \emph{intra-experiment} variability.  
We focus on Adaptive Video Streaming over HTTP. Indeed, video streaming is nowadays widespread and represents the majority of Internet traffic. 
Among tens of protocols dedicated to video streaming, in the last years, most of the providers converged towards the use of video streaming over HTTP that relies on conventional HTTP servers with no specific support. The most common solutions are called Dynamic Adaptive Streaming over HTTP (DASH) and HTTP Live Streaming (HLS), and both realize adaptive bitrate streaming splitting the video in chunks of few seconds. With the growth of high-resolution and 4K video providers, the network is required to ensure high and stable bandwidth in order to guarantee users' QoE. Thus, it is not surprising that video providers publish online rankings of ISP quality.\footnote{For example NetFlix publishes \url{https://ispspeedindex.netflix.com/}}

The goal of this section is to show that \TOOL helps in obtaining realistic figures about video streaming QoE under mobile networks, where the network quality may change while reproducing the video. To this end, we set up a simple DASH server that we use to replay a $200$ seconds full HD (1080p) video. Using Google Chrome instrumented with Selenium Web Driver, we play the video while simultaneously using \TOOL to emulate mobile networks. We perform several experiments, configuring \TOOL to periodically change the network conditions every $n=10$ seconds to emulate a variable network within the selected model. Like in the previous sections, we also perform experiments with the \SIMPLE that we use as a baseline.


The results show the impact of network conditions and their variability on the users' QoE. Indeed, network conditions impact the bitrate at which the video is consumed. Moreover, a variable network leads to frequent adaptions in video bitrate, as the client needs to adjust it to avoid running out of video buffer. These two factors are commonly accepted as objective metrics to study users' QoE~\cite{seufert2014survey,hossfeld2014assessing,juluri2015measurement}. 
Using \TOOL, one can understand how mobile networks impact such metrics and run new experiments to evaluate the impact of different configurations of the player or the server. 
Figure~\ref{fig:video} shows two examples of video sessions for bad 3G and 4G profiles. We focus on profiles with low signal quality, as they are the most challenging for delivering high-quality videos. 
Figure~\ref{fig:video} shows the timeline of the bitrate obtained by the client over the $200$ seconds of the experiment. 
In both cases, with \TOOL (green line), the bitrate varies several times (more than $15$ in both cases), to follow the variability in network conditions. On the contrary, the \SIMPLE (blue line) does not trigger frequent bitrate adaptations, being stuck on 720p most of the time. This is particularly important, as many bitrate adaptations are known to degrade perceived QoE.

In summary, using \TOOL, it is possible to emulate how adaptive video streaming protocols such as DASH and HLS behave under mobile networks. As such, it allows one to understand the impact of network conditions on the video-streaming objective metrics that affect users' QoE.
\section{Conclusion}
\label{sec:conclusions}

Mobile devices have become one of the main tools used for accessing the Internet and web services. However, the heterogeneity of operator deployments, technologies, and mobility makes it highly dynamic and making the prediction of network performance a challenging task. As such, network emulation is of fundamental importance for testing new applications, new protocols, or performing research experiments. 

In this paper, we proposed \TOOL, a simple tool to achieve realistic network emulation. We used the results of a large-scale campaign of speed test measurements to build models representing the network conditions observed in particular scenarios.
The dataset was collected in $2$ countries from $4$ mobile network operators. We used it to implement $32$ distinct networking models for \TOOL.  Indeed, each model is built using all measurements observed for a specific operator, RAT, and signal quality. Using the KDE technique, we estimate the multivariate probability function of latency, download and upload bandwidth. As such, the estimated distributions describe both (i) the \emph{typical} network conditions observed, and (ii) their \emph{variability}. \TOOL uses these distributions to sample new plausible values of network parameters at run-time, and enforces them using the \texttt{tc-netem} Linux tool. Moreover, we provide a guideline to generate new emulation models for \TOOL.

We validated \TOOL using experiments made on operational networks in which we downloaded large-sized HTTP objects. Not only \TOOL leads to similar results in terms of average behavior, but it can recreate the same variability in download time observed in the real networks. Then, we compared \TOOL performance with other freeware network emulation tools showing how we can better recreate the network variability. Moreover, we described two practical use cases for \TOOL. First, we analyzed the performance of web pages under different network conditions. We pointed out significant differences in page load time when varying operator, RAT, or signal quality. Second, we showed how video streaming performance varies in terms of average bitrate and quality-level adjustments when having variable network conditions.

As future work, we plan to gather additional measurements to build more \TOOL models including the 5G technology and mobility scenarios. Moreover, we plan to enrich the complexity of our models to include not only bandwidth and latency, but also other parameters such as jitter, packet loss, and reordering. 

\section*{Acknowledgements}
The research leading to these results has been funded by the European Union’s Horizon 2020 research and innovation program under grant agreement No. 644399 (\monroe) and the Smart-Data@PoliTO center for Big Data technologies.

\biboptions{sort&compress}

\small
\bibliographystyle{ieeetr}
\bibliography{reference}

\end{document}